\documentclass[12pt]{article}

\usepackage{scicite}

\usepackage{times}
\usepackage{graphicx}

\newcommand{\km}{${\rm km\,s}^{-1}$}
\newcommand\nodata{ ~$\cdots$~ }%
\newcommand{\hst}{{\em HST}}

\newcommand{\vlsr}{$v_{\rm LSR}$\relax}

\def\lesssim{\mathrel{\hbox{\rlap{\hbox{%
 \lower4pt\hbox{$\sim$}}}\hbox{$<$}}}}
\def\gtrsim{\mathrel{\hbox{\rlap{\hbox{%
 \lower4pt\hbox{$\sim$}}}\hbox{$>$}}}}
\let\la=\lesssim                
\let\ga=\gtrsim

\newcommand\arcsec{\mbox{$^{\prime\prime}$}}%

\newcommand{\hit}{H$\;${\scriptsize \rm I}\relax}
\newcommand{\oit}{O$\;${\scriptsize \rm I}\relax}
\newcommand{\siiit}{Si$\;${\scriptsize \rm II}\relax}

\newcommand{\hi}{H$\;${\small\rm I}\relax}
\newcommand{\hii}{H$\;${\small\rm II}\relax}

\newcommand{\alii}{Al$\;${\small\rm II}\relax}

\newcommand{\cii}{C$\;${\small\rm II}\relax}
\newcommand{\ciii}{C$\;${\small\rm III}\relax}
\newcommand{\civ}{C$\;${\small\rm IV}\relax}

\newcommand{\oi}{O$\;${\small\rm I}\relax}

\newcommand{\ovi}{O$\;${\small\rm VI}\relax}

\newcommand{\sii}{S$\;${\small\rm II}\relax}
\newcommand{\siii}{Si$\;${\small\rm II}\relax}
\newcommand{\siiii}{Si$\;${\small\rm III}\relax}
\newcommand{\siiv}{Si$\;${\small\rm IV}\relax}

\newcommand{\feii}{Fe$\;${\small\rm II}\relax}

\newenvironment{sciabstract}{%
\begin{quote} \bf}
{\end{quote}}

\newcounter{lastnote}

\title{A Reservoir of Ionized Gas in the Galactic Halo to Sustain Star Formation in the Milky Way}

\author
{Nicolas Lehner$^{\ast}$, J. Christopher Howk \\
\\
\normalsize{Department of Physics, University of Notre Dame,}\\
\normalsize{225 Nieuwland Science Hall, Notre Dame, IN 46556, USA}\\ \\
\normalsize{$^\ast$To whom correspondence should be addressed; E-mail:  nlehner@nd.edu.} \\
\\
\normalsize{Accepted by Science, 25 August 2011 -- To appear in Science, November 18, 2011.
}
}

\date{}

\begin{document} 




\maketitle 

\begin{sciabstract}
Without a source of new gas, our Galaxy would exhaust its supply of gas through the formation of stars.  Ionized gas clouds observed at high velocity may be a reservoir of such gas, but their distances are key for placing them in the Galactic halo and unraveling their role. We have used the Hubble Space Telescope to blindly search for ionized high-velocity clouds (iHVCs) in the foreground of Galactic stars. We show that iHVCs with $90 \le |v_{\rm LSR}| \la 170$ \km\ are within one Galactic radius of the sun and have enough mass to maintain star formation, while iHVCs with $|v_{\rm LSR}| \ga 170$ \km\  are at larger distances. These may be the next wave of infalling material.
\end{sciabstract}

The timescale for gas consumption via star formation in spiral galaxies is far shorter than a Hubble time ($13.8$ billion years), requiring an on-going replenishment of the gaseous fuel in the disks of galaxies for continued star formation. Analytical models and hydrodynamical simulations have emphasized the importance of cold stream accretion as a means for metal-poor gas (metallicity that is less than  $10\%$ solar or  $Z \la 0.1 Z_\odot$)  to flow onto galaxies along dense intergalactic filaments \cite{keres09}. However, galaxies may also exchange mass with the local intergalactic medium (IGM) through outflows driven by galactic ``feedback'',  galactic winds powered by massive stars and their death and from massive black holes.  Some of this material may return to the central galaxy as recycled infalling matter -- the galactic fountain mechanism \cite{fraternali08,oppenheimer10}.  The circumgalactic medium about a galaxy is thus a complicated blend of outflowing metal-rich and infalling metal-poor gas. The relative importance of these processes is poorly constrained observationally. Here we demonstrate that ionized gas in the local Galactic halo provides a major supply of matter for fueling ongoing star formation.

Inflow and outflow in the Milky Way halo can be studied via the so-called high-velocity clouds (HVCs), clouds  moving in the local standard of rest (LSR) frame at $|\mbox{\vlsr}| \ge 90$ \km\ \cite{wakker97}. Determining the distance ($d$) of these HVCs is critical for associating HVCs with flows occurring near the Milky Way rather than the IGM of the Local Group and for quantifying their basic physical properties because several of these directly scale with the distance (e.g., the mass $M\propto d^2$). Major progress has been made in the last decade for some of the large, predominantly neutral HVC complexes, placing them  4--13 kpc from the sun (excluding here the Magellanic Stream) \cite{wakker01,wakker07,wakker08,thom08}. About 37\% of the Galactic sky is covered by \hi\ HVCs with column densities $N($\hi$)\ge 10^{17.9}$ cm$^{-2}$ \cite{murphy95}, but the infall rate of the largest \hi\ HVC complex within 10 kpc (complex C), $\sim 0.14$\,${\rm M}_\odot\,{\rm yr}^{-1}$ \cite{thom08},  is far too modest for replenishing the $0.6$--$1.45$ M$_\odot$\,yr$^{-1}$ consumed by Milky Way star formation \cite{robitaille10}. This is not entirely surprising given that recent models show that inflowing gas should be predominantly ionized in view of the small amounts of neutral gas available for inflow at any epoch \cite{bauermeister10}. In ionized gas, $N($\hi$)$ becomes small relative to $N($\hii$)$, and the \hi\ emission becomes extremely difficult to impossible to detect.  Low \hi\ content HVCs are, however, routinely found in absorption in the spectra of cosmologically distant objects, such as active galactic nuclei (AGNs), with a detection rate of $\sim$60--80\% \cite{sembach03,fox06,collins09,shull09,richter09}. Their total (neutral and ionized) hydrogen column density is shown to be quite large (as large as $N($\hi$)$ in predominantly neutral HVCs). These are therefore ionized HVCs (iHVCs), i.e., $N({\mbox \hii}) \gg N({\mbox \hi})$ \cite{sembach03,shull09}. Given their large covering factor, the iHVCs may represent the long sought supply of gas needed for continued Milky Way star formation.  However, the iHVCs have been mostly detected against AGNs: they may reside within the Galaxy, the Local Group, or the IGM. Thus, as for their larger \hi\ column density counterparts, direct distance constraints are required for determining their masses and for characterizing their role in the evolution of the Milky Way. 

To determine the distances of iHVCs can be directly undertaken by observing the gas in the foreground of stars at known distances from the sun. Recently, based on observations of high-velocity interstellar absorption in the ultraviolet spectra of two Galactic stars, two of these iHVCs were found within 8--15 kpc from the sun \cite{zech08,lehner10}. One of them was found toward the inner Galaxy and has super-solar metallicity, probing gas that has been ejected  from and is raining back onto the Milky Way disk \cite{zech08}. Another was observed at $l\sim 103^\circ$ with a galactocentric distance of $<17.7$ kpc and sub-solar  metallicity; it likely traces gas that is being accreted by the Milky Way, possibly associated with the well known large \hi\ complex C \cite{lehner10,tripp11}. These studies show that some iHVCs probe gas flows in and out of the Galactic disk, with some originating from the Milky Way and others having an extragalactic origin.  

Here we generalize the result to understand these iHVCs in the context of the Milky Way evolution with a survey of the gas in the foreground of $28$ distant Galactic halo stars with known distances. These stars were observed with the Cosmic Origins Spectrograph (COS) and Space-Telescope Imaging Spectrograph (STIS) on board the {\it Hubble Space Telescope (HST)}, and a large majority ($23/28$) of these were obtained through our {\it HST}\ Cycle 17 program 11592 (SOM).  The main criterion for assembling this stellar sample is that these UV bright stars are at height $|z|\ga 3 $ kpc from the Galactic plane. This minimum height was adopted in view of absence or scarcity of iHVCs at smaller $z$ \cite{zsargo03} and other works on their predominantly neutral counterparts \cite{wakker01}. We systematically searched for high-velocity interstellar metal-line absorption in the COS and STIS UV spectra of these stars (SOM). We noted HVC detections only if high velocity absorption was seen in multiple ions or transitions (at a minimum two, see Fig.~1). We measured the apparent optical depth-weighted mean velocity  $\langle v\rangle = \int v \tau_a(v)dv /\int \tau_a(v)dv $ (Table~S1). Most of the HVCs seen in absorption against the stars do not have \hi\ $21$-cm emission at a level of $\ga 10^{18.5}$ cm$^{-2}$ based on the Leiden/Argentine/Bonn (LAB) survey \cite{kalberla05}. The column densities of \siii\ and \oi\ imply large ionization fractions in several HVCs of our sample (Fig.~1, SOM, and  \cite{lehner01,zech08}), demonstrating that $N({\mbox \hii}) \gg N({\mbox \hi})$.

The iHVC detection rate in our stellar sample is $50\%$ ($14/28$). While a sightline may have more than one high-velocity absorption component, only one HVC for a given sightline is counted for estimating the covering factor. Defining a sample with a more uniform sensitivity $W_\lambda \ge 15$ m\AA\ near \siii\ $\lambda$1526 and with no stellar contamination (i.e., with no observed stellar photospheric absorption at $|v_{\rm LSR}| \ge 90$ km; see flag $Q=1$ in Table~S1 and SOM for more details) gives essentially the same detection rate with $47\%$ ($9/19$). The average  distance of the stars in the latter sample where the iHVCs are detected is $11.5 \pm 4.1$ kpc; the average absolute $z$-height is $7.3 \pm 3.0$ kpc; for the whole sample these are $\langle d\rangle = 11.6 \pm 6.9$ kpc and $\langle |z|\rangle = 6.5 \pm 3.2$ kpc. Our sky coverage is larger at $b>0^\circ$ than at $b<0^\circ$ (Fig.~2); the difference of detection rates between the northern ($60\%$, $12/20$) and southern ($29\%$, $2/7$) sky may be due to statistical fluctuations in the smaller southern sample. With the better sampled northern Galactic sky, there is some evidence for an increase in the detection rate to $73\%$ ($11/15$) for sightlines with $z\ga 4$ kpc, implying that most of the iHVCs could be situated at $4\la z \la 9$ kpc.  There is no iHVC absorption at $|v_{\rm LSR} |\ga 170$ \km\ (VHVCs) toward the stars (Fig.~2, SOM) even though these are observed along the path to AGNs (Fig.~2, SOM, and see below). Thus there is no evidence for  VHVCs in the Galactic halo at $|z|\la 10$ kpc. 

In order to understand the implications of the iHVC detection rate toward stars, we need to compare it to a distance-independent measure of the iHVC covering fraction. Earlier studies determined the covering factors of these iHVCs toward AGNs \cite{sembach03,fox06,collins09,shull09}, but they concentrated on a single species (\ovi\ or \siiii), which differs from our search method. We combined observations from \cite{richter09} and \cite{collins09} to assemble an AGN sample using the same search criteria adopted for our stellar survey (i.e., using the same metal ions) (SOM). Our AGN sample is summarized in Table~S2 and its sky distribution is shown on Fig.~2. It has a similar sensitivity and size as the stellar sample.  The detection rate is $67\%$ ($16/24$) for the iHVCs and $42\%$ ($11/26$) for the VHVCs (excluding the $4$ Magellanic Stream sightlines).  For the HVC sample, $6$ sightlines with \hi\ LAB emission at $90 \le |v_{\rm LSR}|\le 170$ \km\ were excluded because of strong selection effects in the AGN sample: many of these AGNs were indeed initially targeted to study  known \hi\ HVCs (including these sightlines yields $f_c=73\%$). After removing those, the AGN sample may still be biased and overestimate the true covering factors because a given AGN could have been specifically targeted for studying an HVC or could have been favored over other AGNs because of previously known HVCs near the line of sight.

The covering factors of the iHVCs are therefore $f_c = 50$\% ($14/28$) and $\le 67$\% ($16/24$)  for the stellar and AGN samples, respectively. Hence a majority (if not all) of the iHVCs seen toward AGNs are within  $\langle d\rangle \simeq 12 \pm 4$ kpc and $\langle |z|\rangle \simeq 6 \pm 3$ kpc, implying that the iHVCs at $90 \la |v_{\rm LSR}| \le 170$ \km\ mostly trace flows of ionized gas in the Milky Way halo. On the other hand, VHVCs must  then lie beyond $d\ga 10$--$20$ kpc  ($|z|\ga 6$--$10$ kpc) because they are not detected toward any star. The VHVCs could be associated with the outer reaches of the Milky Way or gas in the Local Group. In the former scenario, this would imply that iHVCs slow through their interaction with the Galactic gas as they approach the plane, as predicted by some models of HVCs \cite{benjamin97,peek07,heitsch09, marinacci11}.  In the latter scenario, some of the VHVCs could be at large distances from the Milky Way or other galaxies (e.g., Andromeda) and be a component of the multiphase local intergalactic medium, which may permeate the Local Group of galaxies \cite{blitz99,nicastro03}.  

We note that a majority of the so-called \ovi\ HVCs have accompanying \ciii\ and \hi\ HVC absorption at similar high velocities, demonstrating their multiphase nature \cite{fox06}. The velocity sky-distribution of the \ovi, \siiii, and \hi\ HVCs are also alike \cite{sembach03,fox06,collins09}, and the sky-distribution of the iHVCs seen toward the AGNs and stars is moreover remarkably similar considering our better sampled north Galactic sky (Fig.~2). These and the fact that the \hi\ HVCs and iHVCs are now known to be at similar distances strongly suggest they are all related, probing separate phases where the \hi\ HVCs may be the densest regions of the more diffuse ionized complexes. 

Having demonstrated the iHVCs are in the local Galactic halo, we can reliably estimate their mass and assess their importance for future star formation in the Milky Way. The mass of these iHVCs can be estimated $M_{\rm iHVC} \simeq 1.3 m_{\rm H} (4\pi d^2) f_c N_{\rm H\,II} \approx 9.6 \times 10^{-12} (d/{\rm 12\, kpc})^2 (f_c/0.5)N_{\rm HII}$ M$_\odot$ (where $m_{\rm H}$ is the mass of hydrogen and the factor $1.3$ accounts for the additional mass of helium). The ionized gas probed by \siii, \siiii, and \siiv\ has $\langle N_{\rm H\,II}\rangle \approx 6\times 10^{18} (Z/0.2Z_\odot)^{-1}$ cm$^{-2}$ \cite{shull09} while the \ovi\ phase has $\langle N_{\rm H\,II} \rangle \approx 4.5\times 10^{18} (Z/0.2Z_\odot)^{-1}$ cm$^{-2}$ \cite{sembach03,fox06,shull09}, where the metallicity of $0.2 Z_\odot$ is representative of complex C and other large \hi\ complexes \cite{wakker01,tripp11}. As \ovi\ and \siiii\ are not expected to exist in the same gas phase, the total \hii\ column density is a sum of the phases traced by these ions, $N_{\rm H\,II}\approx 1.1 \times 10^{19} (Z/0.2Z_\odot)^{-1}$ cm$^{-2}$. These assumptions imply a total mass $M_{\rm iHVC} \approx 1.1 \times 10^8 (d/{\rm 12\,kpc})^2 (f_c/0.5)(Z/0.2Z_\odot)^{-1}$ M$_\odot$. We estimate infall times of 80--130 Myr assuming infall velocity of about $90$--$150$ \km,  which implies a mass infall rate for the ionized gas of $\sim 0.8$--$1.4$ M$_\odot$\,yr$^{-1}$ ($d=12$ kpc, $f_c=0.5$, and $Z = 0.2 Z_\odot$). While this value may be reduced somewhat because there must be a mixture of outflows and inflows, large \hi\ complexes have subsolar abundances \cite{wakker01}, suggesting a substantial fraction of the circumgalactic neutral and ionized gas is infalling. Some of the outflowing gas must also be recycled via a Galactic fountain as the iHVCs do not reach the Milky Way escape velocity. We  did not bracket the distances of the iHVCs, but their distances are unlikely to be much smaller than $3$ kpc as otherwise they would already have been detected serendipitously in the UV spectra of more nearby stars \cite{zsargo03}. If all the iHVCs were at $d=5$ kpc, the infall rate would still be significant with $\sim 0.4$--$0.6$ M$_\odot$\,yr$^{-1}$. 

Although the infall rate  depends on several parameters that are not well constrained, our distance estimates allow us to unambiguously place the iHVCs at  $90 \le |v_{\rm LSR}| \le 170$ \km\ in the Milky Way's halo. Assuming an average metallicity of $0.2 Z_\odot$, the estimated infall rate is a factor $6$--$10$ larger than that of the large \hi\ complex C \cite{wakker07,thom08}. (Even with an unrealistically solar metallicity value for all the iHVCs, their infall rate would still be important.) The present-day total star formation rate of the Milky Way is $0.6$--$1.45$ M$_\odot$\,yr$^{-1}$ \cite{robitaille10}, and chemical evolution models require a present-day infall rate of only $0.45$ M$_\odot$\,yr$^{-1}$ \cite{chiappini01}, a value that could be somewhat reduced according to stellar mass loss models \cite{leitner11}. The iHVCs are thus sufficient to sustain star formation in the Milky Way.

We have implicitly assumed the HVCs survive their fall onto the Galactic disk. Hydrodynamical simulations of HVCs support this idea \cite{heitsch09,marinacci11}.  In these models, the HVCs lose most of their \hi\ content within $10$ kpc of the disk and  continue to fall toward the disk as warm ionized matter. This is consistent with the iHVCs covering more Galactic sky and tracing a larger mass reservoir than the predominantly neutral HVCs.  As the iHVCs are still overdense relative to the halo medium, they can continue to sink toward the Galactic plane where they decelerate and feed the warm ionized medium (Reynolds layer) $1$--$2$ kpc from the Galactic disk. In the Reynolds layer, the clouds have low velocities and cannot be identified as HVCs anymore, but as low- or intermediate-velocity clouds, consistent with the observed absence of iHVCs at low $z$-height.  In this scenario, VHVCs are also not expected to be seen near the Galactic disk as the HVC velocities decrease with decreasing $z$-height due to drag. This is  consistent with the lack of VHVCs at $|z| \la 10$--$20$ kpc and the observed velocity sky-distribution \cite{benjamin97}.

\bibliographystyle{Science}


\noindent
Acknowledgments. Based on observations made with the NASA/ESA Hubble Space Telescope, obtained at the Space Telescope Science Institute (STScI), which is operated by the Association of Universities for Research in Astronomy, Inc. under NASA contract No. NAS5-26555. We greatly appreciate funding support from NASA  grant HST-GO-11592.01-A from STScI. We are grateful to Dr. Pierre Chayer for computing several stellar spectra for us. The data reported in this paper are tabulated in the SOM and archived at the Multimission Archive at STScI (MAST, http://archive.stsci.edu/).



 \begin{figure}
 \includegraphics[width = 6 truecm]{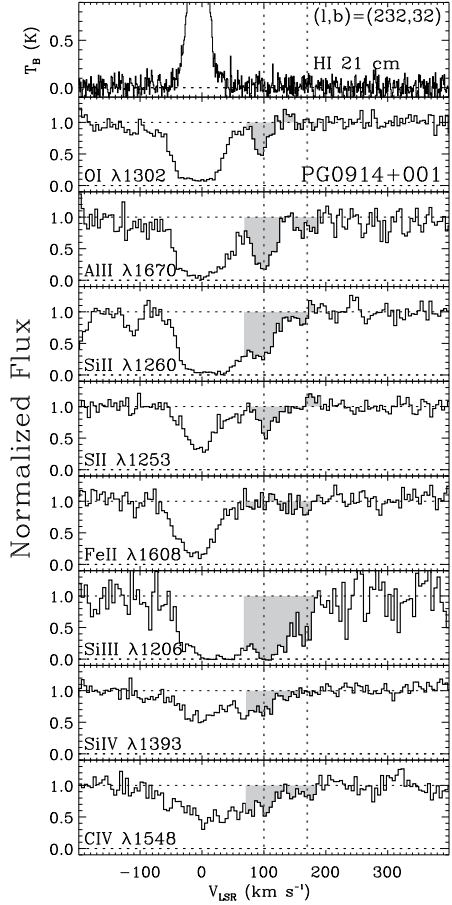}
 \caption{\small  Example of COS continuum normalized absorption profiles  of various metal-lines and LAB H\,I emission line profile ({\it top panel}) toward PG0914$+$001, a star at $d=16$ kpc and $z=+8.4$ kpc. There are at least two iHVCs seen in absorption at $+100$ and $+170$ \km\ indicated by the dotted lines and shaded regions. The $\sim 0$ \km\ absorption and H\,I emission are from the Milky Way disk. We derive $[${\rm O\,I/S\,II}$] \equiv \log(N({\rm O\,I})/N({\rm S\,II})) - \log(A_{\rm O}/A_{\rm S})_\odot \simeq -2.34$, implying H\,II\,$\gg$\,H\,I. If the iHVC has a solar abundance, $N($H\,II$)>10^{19.6}$ cm$^{-2}$ (based on S\,II) and  $N($H\,I$)\simeq 10^{17.30}$ cm$^{-2}$ (based on O\,I). These column densities would increase if the iHVC metallicity is subsolar. We find  very subsolar $[${\rm Fe\,II/S\,II}$]<-1.5$ and $[${\rm Si\,II/S\,II}$]\simeq -1.4$ ratios, indicating strong ionization and dust depletion effects; the latter could suggest a Galactic origin. Examples of negative-velocity iHVCs toward stars can be found in \cite{zech08,lehner10}.  
  \label{f-spec}}
  \end{figure}

  \begin{figure}
  \includegraphics[width = 16truecm]{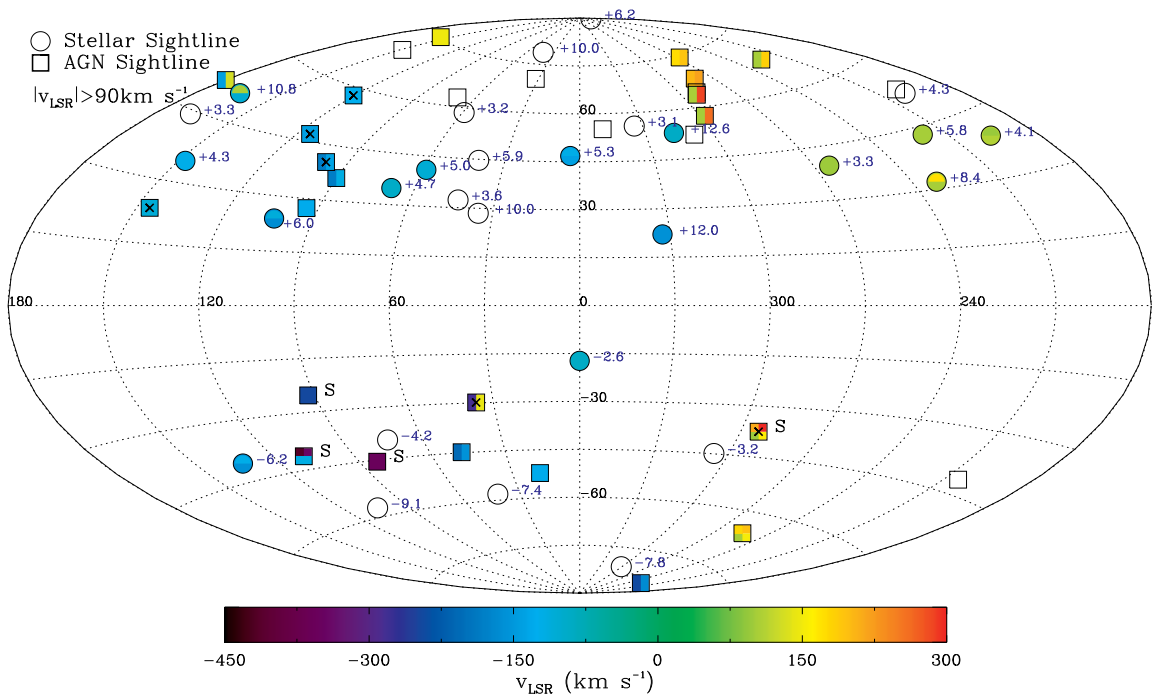}
  \caption{\small Aitoff projection Galactic (longitude, latitude) map of the survey directions for the stellar sample (circles) and extragalactic sample (square). A colored circle/square indicates an HVC in the foreground of the star/AGN while blank circle/square implies no HVC along the stellar/AGN sightline. The velocity value is color coded following the horizontal color bar.  The positive and negative numbers indicate the $z$-height (in kpc) of the stars. The squares with a X are HVCs where H\,I\ $21$-cm LAB emission is present at similar velocities seen in absorption. A S near a square indicates a sightline that passes through the Magellanic Stream.\label{f-map}}
  \end{figure}

\newpage
\begin{center}
\section*{Supporting Online Material}
{\Large \bf A Reservoir of Ionized Gas in the Galactic Halo to Sustain Star Formation in the Milky Way}
\\
\vspace{0.3 cm}
{\Large Nicolas Lehner, J. Christopher Howk} \\
\vspace{0.3 cm}
\normalsize{Department of Physics, University of Notre Dame,}\\
\normalsize{225 Nieuwland Science Hall, Notre Dame, IN 46556, USA}\\
\end{center}

\paragraph*{Observations and Analysis.}
Our COS and STIS program consists of $23$ early-type and post-asymptotic giant branch (PAGB) field stars as well as globular cluster (GC) stars  obtained from our Cycle $17$ \hst\ program (\#$11592$). The criteria for assembling this stellar sample was that the stars are at $|z|\ga 3 $ kpc and are bright enough to be observed at a rate of $1$ \hst\ orbit per star. We also searched the literature and the MAST archive for any additional stars at $|z|\ga 3 $ kpc observed with STIS or COS, which added $5$  stars to the sample. (We note that although BD+38$^\circ$2182 is at $z=+3.2$ kpc, this sight line was initially targeted for the HVC Complex M, and hence is not included here.)  In Table~S\ref{t-star}, we list the distance and positions of the $28$ stars.  Stars associated with GCs based on their location and velocity were placed at distances of the hosting GC. For all the field stars, distances are spectroscopic parallaxes  based on models of stellar atmospheres, which produce stellar distances accurate to $\sim$\,$20\%$. The distances were derived by \cite{ramspeck01} and \cite{smoker03} and references therein. 

Sixteen stars were observed with COS with gratings G130M and G160M (resolution $R\simeq 17,000$), and the 5 brightest stars of our sample were obtained with STIS with the echelle E140M mode ($R\simeq 45,800$) using the $0.2\arcsec$ square aperture to maximize throughput. Although the COS resolution is $\sim 2.7$ lower than the STIS resolution, it was adequate for searching for and detecting HVC absorption since the high-velocity absorption is sufficiently separated from the lower velocity Galactic gas. The wavelength ranges covered by COS and STIS observations are $1134$ to $1796$ \AA\ and $1170$ to $1730$ \AA, respectively. However, the three faintest stars (PG1002+506, PG0914+001, and NGC104-UIT14) were only observed with the  COS G160M giving wavelength coverage $1382\le \lambda \le 1747$ \AA. In these wavelength intervals, some or all of the following atomic and ionic species are available: \oi\ $\lambda$1302, \cii\ $\lambda$$1334$, \civ\ $\lambda\lambda$$1548, 1550$, \sii\ $\lambda\lambda$$1250, 1253, 1259$, \siii\ $\lambda$$1190, 1193, 1260, 1304, 1526$, \siiii\ $\lambda$$1206$, \siiv\ $\lambda\lambda$$1393, 1402$, \alii\ $\lambda$$1670$, \feii\ $\lambda$1608. All these atomic and ionic species were used to search high-velocity absorption components in the COS and STIS data. 

Information about the COS instrument and COS spectra can be found in \cite{froning09,dixon10}. Information about STIS can be found in the STIS {\it HST}\ Instrument Handbook \cite{dressel07}. Standard reduction and calibration procedures were performed using the reference files available as of October 2010 or later. The proper alignment of the individual spectra was achieved through a cross-correlation technique. Data from individual exposures, including all grating configurations for COS and echelle orders for STIS, were combined to produce a single spectrum. 

\begin{table}[!t] 
\begin{minipage}{15 truecm}
\caption{Stellar Sample \label{t-star}}
\footnotesize
\begin{tabular}{lccccccccc}
\hline
\hline
{Name} & {$l$}& {$b$} & 
$d$ & $z$ & $v \sin i$& $v^\star_{\rm LSR}$& $v^{\rm HVC}_{\rm LSR}$ &$Q$ & Note\\
 & ($^\circ$)& ($^\circ$) &  (kpc)& (kpc) & (\km) & (\km) & (\km)  &  & \\
(1) & (2)& (3) &  (4)& (5) & (6) & (7) & (8)  & (9)  & \\

\hline
NGC6723-III60	 &     0      &    $   -17     $    &	   8.7     &	$     -2.6  $ &     \nodata  &   $   -101  $   &   $  -90		  $	&    0	    &     \\  
NGC5904-ZNG1	 &     4      &    $   +47     $    &	   7.2     &	$     +5.3  $ &     \nodata  &   $   +65   $   &   $  -140,-120 	  $	&    1      & $a$      \\  
PG1708+142	 &     35     &    $   +29     $    &	 21.0	   &	$  +10.0    $ &     \nodata  &   $    -28  $   &      No			&    1      &   \\  
PG1704+222	 &     43     &    $   +32     $    &	   6.9     &	$  +3.6     $ &     \nodata  &   $   -22   $   &      No			&    1      &   \\  
PG1610+239  	 &     41     &    $   +45     $    &	   8.4     &	$  +5.9     $ &       75     &   $   +109   $   &     No\,$:$			&    0      &   \\  
PHL346  	 &     41     &    $   -58     $    &	   8.7     &	$  -7.4     $ &       45     &   $   +66   $   &     No 			&    0      &   \\  
vZ\,1128    	 &     43     &    $	+79    $     &      10.2    &	 $  +10.0    $ &     \nodata  &   $ -137    $	&     No		         &   0      & $a$   \\  
NGC6205-Barnard29&     59     &    $	+41    $     &      7.7     &	 $  +5.04    $ &     \nodata  &   $ -227     $	&   $  -107		   $     &    1     & $a$   \\
PG1511+367	 &     59     &    $   +59     $    &	   3.8     &	$  +3.2     $ &       75     &   $  +118   $   &     No 			&     1     &   \\  
NGC6341-326	 &     68     &    $   +35     $    &	   8.2     &	$  +4.7     $ &     \nodata  &   $  -101   $   &   $  -94:		  $	&    0      &   \\  
PG2219+094	 &     73     &    $   -40     $    &	   6.6     &	$  -4.2     $ &      225     &   $  -17    $   &     No 			&    1      &   \\  
HS1914+7139	 &    103     &     $  +24	$   &	 14.9	  &	$  +6.0     $ &      250     &   $  -25    $   &   $ -165,-112$ 		&    1      & $b,c$  \\  
PG0009+036	 &    105     &     $  -58	$   &	 10.8	  &	$  -9.1     $ &      350     &   $  +150   $   &      No			&    1      &   \\  
PG0122+214	 &    133     &     $  -41	$   &	  9.6	  &	$  -6.2     $ &      117     &   $  +24    $   &   $ -160,-91	  $		&      1    &   \\  
PG0832+675 	 &    148     &     $  +35	$   &	  7.5	  &	$  +4.3     $ &     \nodata  &   $  -70    $   &   $  -123		  $	&      1    & $b,d$  \\  
PG1002+506	 &    165     &     $  +51	$   &	  13.9    &	$  +10.8    $ &      350     &   $    0    $   &   $  -102,+101 	  $	&    1      &   \\  
HD233622	 &    168     &     $  +44 	$    &     4.7     &	 $  +3.3     $ &    240       &   $ +9      $	&      No		         &    1     & $a$      \\  
PG0855+294	 &    196     &     $ +39	$   &	  6.5	  &	$  +4.1     $ &      100     &   $  +73   $    &   $   +93,+107		 $	&    1      &    \\
PG0955+291	 &    200     &     $  +52	$   &	  5.5	  &	$  +4.3     $ &      190     &   $   +72   $   &      No			&    1      &   \\  
PG1243+275	 &    207     &     $  +89	$   &	  6.2	  &	$  +6.2     $ &     \nodata  &   $  +107   $   &   $  (+105)  	 	$	&    0      & $e$  \\  
PG0934+145	 &    219     &     $  +43	$   &	  8.5	  &	$ +5.8      $ &      50:     &   $  +105   $   &   $  (+100)  	  	$	&     0     & $f$  \\  
PG0914+001       &    232     &     $  +32 	 $   &    16.0     &	 $ +8.4      $ &      350     &   $  +80    $	&   $  +100,+170 	   $     &   1      &    \\  
EC10500-1358     &    264     &     $  +40 	 $   &     5.2     &	 $  +3.3     $ &      100     &   $  +84    $	&   $  +97		   $     &   0      &    \\  
SB357            &    301     &     $  -81 	 $   &     7.9     &	 $  -7.8     $ &      180     &   $  +52    $	&      No		         &    1     &     \\  
NGC104-UIT14     &    306     &     $  -45 	 $   &     4.5     &	 $  -3.2     $ &     \nodata  &   $  +55    $	&      No		         &    1     &  $b,g$   \\  
PG1323-086       &    317     &     $  +53 	 $   &    15.8     &	 $ +12.6     $ &     \nodata  &   $  -41    $	&   $  -91		   $     &    1     &    \\  
NGC5824-ZNG1     &    333     &     $  +22 	 $   &     32.0    &	 $  +12.0    $ &     \nodata  &   $  -20    $	&   $  -160:		   $     &     0    &    \\  
HD121968	 &    334     &     $  +56 	$    &     3.8     &	 $  +3.1     $ &    160       &   $ +137     $	&      No		         &    1     &  $a$      \\
\hline
\end{tabular}
(1): Name of the background star. \\
(2) and (3): Galactic longitude and latitude of the star. \\
(4) and (5): Distance and vertical $z$-height of the star. The stellar distances are accurate to $\sim$\,20\% \cite{ramspeck01,smoker03}. \\
(6): Projected rotational velocity of the star. If no value is listed, the star is a PAGB or GC star. \\
(7) and (8): Radial velocity of the star and the HVC if present. 
The velocities are averaged over multiple transitions. 
The uncertainties in absolute velocity scales are  $\sim$10 \km\ for COS and $\le 3$ \km\ for STIS, except where there is a colon. A colon means the result is tentative in view of the low signal-to-noise of the data. \\
(9): Quality flag; $Q=0$ means either low S/N ratio ($< 15$) or a possible stellar contamination. \\
{\sc Notes ---} $a$: Not initially in our {\it HST}\ Cycle 17 program. 
$b$: HVC \hit\ emission is observed. 
$c$: The \hit\ emission seen toward HS1914+7139 at $-118$ \km\  is associated with the Outer Spiral Arm, but there is also a second HVC along this line of sight with no \hit\ emission ({\it 18}).
$d$: For PG0832+675, the HVC velocity in absorption does not coincide with the \hit\ emission velocity, and therefore we assume that they are not directly related. 
$e$: high-velocity absorption is observed but it is very likely stellar. 
$f$: there are possibly both stellar and interstellar high-velocity components at the same velocity. 
$g$: Toward NGC104-UIT14, we did not find any evidence of HVC in absorption while it is seen in emission at about $+150$ \km. This suggests that this HVC seen in \hit\ emission is at $d>4$ kpc ($z<-3.5$ kpc) or that the pencil-like beam through the COS aperture did not cover the emission seen in the much larger \hit\ beam.
\end{minipage}
\end{table}
\normalsize
\clearpage

One concern with the use of stars as background continuum sources is the possible stellar contamination. In particular, the PAGB and evolved GC stars may have narrow atmospheric lines that can mimic interstellar absorption. We minimize this issue by requiring interstellar HVC absorption to be present in multiple transitions and by using synthetic spectra derived from model stellar atmospheres to identify possible contaminating stellar lines. The most metal-rich evolved stars in our samples are  NGC6723-III60, PG0943+145, and PG1243+275. For these stars, model atmospheres  were calculated by Dr. P. Chayer in a manner similar to that described in \cite{sonneborn02} using the publicly available TLUSTY and SYNSPEC codes of Drs. I. Hubeny and T. Lanz (http://nova.astro.umd.edu/). The models used the stellar atmosphere parameters previously derived by \cite{moehler98,rolleston99}. The stellar contamination is, however, less of a concern for the rapidly rotating B-type stars in our sample (see the large projected rotational velocity $v \sin i$ in Table~S\ref{t-star}), except for PHL346, a relatively slowly rotating and extremely metal rich star. Yet, even in that case it is possible to rule out the presence of any HVC in this sightline following our methods. We are therefore confident that the detected HVCs are not stellar, except possibly for two stars (see Table~S\ref{t-star}). We also emphasize that the HVC absorption is found blueshifted and redshifted relative to the star velocity. Thus the HVCs are not circumstellar material (see, also, the ionization model and other arguments in {\it 17}). 

From the whole sample of stars, we also built a sample ($Q=1$, see Table~S\ref{t-star}) in which the S/N ratio is more uniform with S/N\,$\ge 15$, i.e., a sensitivity $W_\lambda \ge 15$ m\AA\ near \siii\ $\lambda$1526. We also require no stellar contamination for the $Q=1$ sample, i.e., any star with some detected narrow photospheric features at $90 \le |v^\star_{\rm LSR}|\la 170$ \km\ that correspond to the same species used to search HVC absorption is rejected from the $Q=1$ sample (e.g., vZ\,1128). 

Finally, the signal-to-noise in the spectra of $3$ stars (PG1610+239, NGC6341-326, and NGC5824-ZNG1) is much below than in the spectra of the other stars. We therefore consider the identification or non-detection of the HVCs along these sightlines tentative and mark those with colons in Table~S\ref{t-star} to emphasize this. 

\paragraph*{Extragalactic Sample.}
We built our extragalactic sample based on earlier works where the background AGNs were observed with {\it HST}/STIS E140M ({\it 14,15}).  By using these works and references therein and analyzing some spectra ourselves, we retrieved or estimated the velocities of the detected HVCs toward AGNs. We summarize the velocities in Table~S\ref{t-extra}, and as in the stellar sample, there are sometimes several HVCs at different velocities along a given sightline. We note that the spectra in the stellar and AGN samples have similar sensitivity, and both samples have a similar size. We also emphasize that as for the stellar sample, we require that a given HVC is detected in absorption in at least two different ions. For this reason we do not confirm all the \siiii\ HVCs listed by ({\it 14}). As for the stellar sample we also searched each sightline for possible HVCs seen in \hi\ $21$-cm emission using the LAB data ({\it 21}); the sightlines with \hi\ emission are listed in Table~S\ref{t-extra}. In Table~S\ref{t-extra}, we also mark the sightlines that pass through the Magellanic Stream.

\begin{table}[!t]
\begin{minipage}{9.7 truecm}
\caption{Extragalactic Sample \label{t-extra}}
\footnotesize
\begin{tabular}{lcccc}
\hline
\hline
Name& $l$& $b$ & $v^{\rm HVC}_{\rm LSR}$ & \hit\ 21 cm \\
 & ($^\circ$) & ($^\circ$)  & (\km) & Emission   \\
(1) & (2)& (3) &  (4)& (5)  \\
\hline
PKS2155-304   &  17  &   $ -52  $&   $ -130	 $  		&  No	   \\ 
   NGC5548    &  31  &   $ +70  $&  	No    			&  No	   \\ 
    Mrk509    &  35  &   $ -29  $&   $ -284,+139 $  		&  Yes     \\ 
PG1444+407    &  69  &   $ +62  $&  	No     			&  No	   \\ 
   PHL1811    &  47  &   $ -44  $&   $ -208,-159	 $  	&  No	   \\ 
   NGC7469    &  83  &   $ -45  $&   $ -355		$  	&  No$^a$  \\ 
     3C351    &  90  &   $ +36  $&   $ -185,-118 $ 		&  No	   \\ 
   UGC12163   &  92  &   $ -25  $&   $ -245		 $  	&  No$^a$  \\ 
 H1821+643    &  94  &   $ +27  $&   $ -130 $  			&  No	   \\ 
    Mrk876    &  98  &   $ +40  $&   $ -185,-137 $  		&  Yes     \\ 
    Mrk335    & 108  &   $ -41  $&   $ -416,-342,-118  $  	&  No$^a$ \\ 
    Mrk279    & 115  &   $ +46  $&   $ -145 $  			&  Yes     \\ 
PG1259+593    & 120  &   $ +58  $&   $ -125 $  			&  Yes     \\ 
HS0624+690    & 145  &   $ +23  $&   $ -108 $  			&  Yes     \\ 
   NGC4051    & 148  &   $ +70  $&  	No     			&  No	   \\ 
   NGC4151    & 155  &   $ +75  $&   $ +145 $  			&  No	   \\ 
PG0953+414    & 179  &   $ +51  $&   $ -145,+125	$	&  No	   \\ 
     TON28    & 200  &   $ +53  $&  	No    			&  No	   \\ 
PKS0405-123   & 204  &   $ -41  $&  	No	    		&  No	   \\ 
PG1116+215    & 223  &   $ +68  $&   $ +100,+180 $  		&  No	   \\ 
   TONS210    & 224  &   $ -83  $&   $ -243,-167 $  		&  No     \\ 
HE0226-4410   & 253  &   $ -65  $&   $ +99,+148,+175,+193 $  	&  No	   \\ 
PG1211+143    & 267  &   $ +74  $&   $ +174,+191 $  		&  No	   \\ 
PG1216+069    & 281  &   $ +68  $&   $ +199,+216 $  		&  No	   \\ 
     3C273    & 289  &   $ +64  $&  	No	     		&  No	   \\ 
PKS0312-77    & 293  &   $ -37  $&   $+97,+160,+210,+295$       &  Yes$^{a,b}$ \\
Q1230+0115    & 291  &   $ +63  $&   $ +103,+284 $  		&  No	   \\ 
   NGC4593    & 297  &   $ +57  $&   $ +98,+258 $  		&  No	   \\ 
PKS1302-102   & 308  &   $ +52  $&  	No	     		&  No	   \\ 
   Mrk1383    & 349  &   $ +55  $&  	No	    		&  No	   \\ 
\hline
\end{tabular}
(1): Name of the background QSO/AGN. \\
(2) and (3): Galactic longitude and latitude of the QSO/AGN. \\
(4): Radial velocity of the HVC if present. \\
(5): LAB \hit\ emission present at similar high velocities seen in absorption. \\
{\sc Notes ---} $a$: Sightline passing through the Magellanic Stream. \\
$b$: Sightline passing through the Magellanic Bridge. 
\end{minipage}
\end{table}
\normalsize

\paragraph*{Ionization Fraction for the HVCs in the Stellar Sample.}
By comparing the column densities of \siii\ and \oi, we can estimate the ionization fraction in the HVCs. Indeed \oi\ is an excellent tracer of \hi\ because \oi\ and \hi\ have nearly identical ionization potentials and are strongly coupled through charge exchange reactions. On the other hand, the ionization potential of \siii\ is larger than that of \hi\ ($13.6$ eV) allowing this ion to be present in both neutral and ionized gas. In Table~S\ref{t-col}, we summarize the column densities of \oi\ and \siii\ and their ratios corrected for the relative solar abundance for the HVCs where the blending with lower velocity gas or stellar lines is not an issue. The column densities were estimated through $N = \int 3.768\times 10^{14} \ln[F_c(v)/F_{\rm obs}(v)]/(f\lambda) dv$ cm$^{-2}$, where $F_c(v)$  and $F_{\rm obs}(v)$ are the modeled continuum flux and observed flux as a function of velocity, respectively, $f$ is the oscillator strength of the considered atomic transition and $\lambda$ (in \AA) its wavelength \cite{savage91}. When \oi\ is not detected, we estimated a $3\sigma$ upper limit of the column density following \cite{lehner08}.  The absorption features are weak enough that saturation is not an issue. For $5$ HVCs, $[{\mbox \oi}/{\mbox \siii}]$ ratios imply ionization fraction \hii/(\hi$+$\hii)\,$\ge 57\%$--$95\%$. We emphasize that we only consider \siii, but other higher ionization stages are known to be present as revealed by the detection of \siiii\ and \siiv. This implies that $[{\mbox \oi}/{\mbox \siii}]$ gives a strict lower limit on the ionization fraction. For the other $5$ HVCs, we could only estimate $3\sigma$ upper limits based on the non-detection of \oi. They are all consistent with a significant ionization fraction. Therefore the HVCs seen in the foreground stars in our sample can be labeled iHVCs.

\begin{table}[!t]
\begin{minipage}{11.5 truecm}
\caption{Ionization Fraction for the HVCs in the Stellar Sample \label{t-col}}
\footnotesize
\begin{tabular}{lcccc}
\hline
\hline
Name & $v^{\rm HVC}_{\rm LSR}$ & $\log N({\mbox \oit})$&  $\log N({\mbox \siiit})$ & $[{\mbox \oit}/{\mbox \siiit}]$$^a$  \\
 	& (\km) & $[{\rm cm}^{-2}]$& $[{\rm cm}^{-2}]$ &  \\
\hline
NGC5904-ZNG1   &    $  -140  $  &   $	13.12 \pm 0.06    $  &   $  13.32 \pm 0.03   $  &   $	 -1.38 \pm 0.07    $      \\
NGC5904-ZNG1   &    $  -120  $  &   $	13.09 \pm 0.06    $  &   $  13.03 \pm 0.03   $  &   $	 -1.12 \pm 0.07    $      \\
HS1914+7139    &    $  -165  $  &   $	13.83 \pm 0.08    $  &   $  13.10 \pm 0.10   $  &   $	 -0.45 \pm 0.13    $      \\
PG0122+214     &    $  -160  $  &   $  <13.44		  $  &   $  12.36 \pm 0.05   $  &   $ <  -0.10  	   $        \\
PG0122+214     &    $	-91  $  &   $	14.17 \pm 0.04    $  &   $  13.36 \pm 0.04   $  &   $	 -0.37 \pm 0.06    $        \\
PG1002+506     &    $  -102  $  &   $  <13.34		  $  &   $  12.47 \pm 0.12   $  &   $ <  -0.31  	   $        \\
PG1002+506     &    $  +101  $  &   $  <13.41		  $  &   $  12.42 \pm 0.15   $  &   $ <  -0.19  	   $        \\
PG0855+294     &    $  +86   $  &   $  <12.98		  $  &   $  12.50 \pm 0.04   $  &   $ <  -0.70  	   $        \\
PG0855+294     &    $  +107  $  &   $  <12.91		  $  &   $  12.64 \pm 0.03   $  &   $ <  -0.91  	   $        \\
PG0914+001     &    $  +100  $  &   $	13.96 \pm 0.04    $  &   $  13.15 \pm 0.03   $  &   $	 -0.36 \pm 0.05    $        \\
\hline
\end{tabular}
{\sc Notes ---} $a$: $[{\mbox \oit}/{\mbox \siiit}] \equiv \log (N({\mbox \oit})/N({\mbox \siiit})) - \log (A_{\rm O}/A_{\rm Si})_\odot) $ where the solar abundances for O and Si are from \cite{asplund09}. Values preceded by ``$<$" are 3 sigma upper limits (\oit\ absorption is not detected). Errors are at the 1$\sigma$ level. 
\end{minipage} 
\end{table}

\end{document}